\documentclass[journal]{IEEEtran}

\usepackage{graphicx}
\usepackage{booktabs}
\usepackage{multirow}
\usepackage{tabularx}
\usepackage{array}
\usepackage{amsmath}
\usepackage{amssymb}
\usepackage{cite}
\usepackage{url}
\usepackage{xcolor}
\usepackage{tikz}
\usetikzlibrary{positioning,arrows.meta,fit,calc,backgrounds,shapes.geometric,mindmap,shadows}
\usepackage{soul}
\usepackage[colorlinks=true,linkcolor=blue,citecolor=blue,urlcolor=blue]{hyperref}
\usepackage{balance}
\usepackage{stfloats}
\usepackage{enumitem}

\renewcommand{\arraystretch}{1.15}
\sethlcolor{yellow}
\soulregister\cite7
\soulregister\ref7
\soulregister\eqref7

\definecolor{cEnergy}{HTML}{E69F00}   
\definecolor{cCarbon}{HTML}{4D4D4D}   
\definecolor{cWater}{HTML}{0072B2}    
\definecolor{cMaterial}{HTML}{009E73} 
\definecolor{cLand}{HTML}{CC79A7}     
\definecolor{cAgent}{HTML}{D55E00}    

\tikzset{
  fig/.style={line width=0.6pt, font=\footnotesize, text=black},
  box/.style={draw=#1!75, fill=#1!12, rounded corners=2pt, line width=0.6pt,
              align=center, inner sep=2.5pt},
  box/.default=cCarbon,
  plainbox/.style={draw=cCarbon!75, fill=white, rounded corners=2pt, line width=0.6pt,
              align=center, inner sep=2.5pt},
  exclbox/.style={draw=cCarbon!75, fill=cCarbon!5, rounded corners=2pt, line width=0.6pt,
              dashed, align=center, inner sep=2.5pt},
  band/.style={draw=#1!75, fill=#1!15, rounded corners=2pt, line width=0.6pt},
  arr/.style={-{Stealth[length=2mm]}, line width=0.6pt, draw=cCarbon},
  link/.style={line width=0.6pt, draw=cCarbon},
  primary/.style={circle, draw=#1!75, fill=#1!75, line width=0.6pt,
              minimum size=2.7mm, inner sep=0pt},
  secondary/.style={circle, draw=#1!75, fill=white, line width=0.9pt,
              minimum size=2.5mm, inner sep=0pt},
  chip/.style={draw=#1!75, fill=#1!15, rounded corners=2pt, line width=0.6pt,
              font=\scriptsize\bfseries, inner sep=1.6pt, minimum width=7.5mm},
}

\begin{document}

\title{Environmental Impact of Generative and Agentic AI: An in-Depth Analysis and Green Solutions}

\author{%
\textbf{Abderaouf Bahi\textsuperscript{1\S}}~\href{https://orcid.org/0009-0003-7116-5080}{\includegraphics[height=1em]{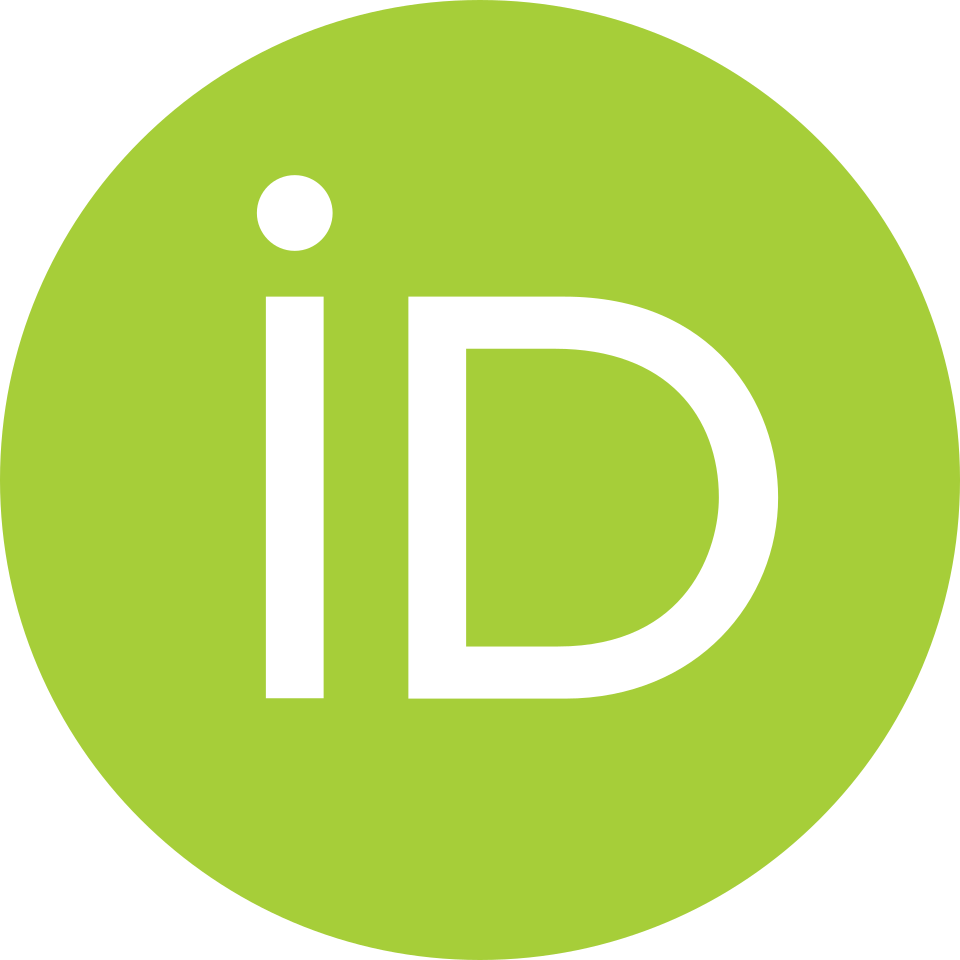}}, 
\textbf{Amel Ourici\textsuperscript{2}}~\href{https://orcid.org/0000-0001-5701-8576}{\includegraphics[height=1em]{ORCID_iD.png}}, 
\textbf{Ibtissem Gasmi\textsuperscript{1}}~\href{https://orcid.org/0000-0002-8939-1727}{\includegraphics[height=1em]{ORCID_iD.png}} \\[0.5em]
\textsuperscript{1}\small
Computer Science and Applied Mathematics Laboratory (LIMA),
Faculty of Science and Technology, Chadli Bendjedid University, 
P.O. Box 73, El Tarf 36000, Algeria \\[0.5em]
\textsuperscript{2}\small
Mathematical Modeling and Numerical Simulation Laboratory (LAM2SIN),
Faculty of Technology, Badji Mokhtar University, 
P.O. Box 12, Annaba 23000, Algeria \\[0.5em]
\textsuperscript{\S}Corresponding author: \textbf{Abderaouf Bahi} (a.bahi@univ-eltarf.dz) \\[0.5em]

}

\maketitle

\begin{abstract}
The proliferation of generative and agentic artificial intelligence (AI) systems has introduced computational demands whose environmental consequences are substantial yet underexamined. This paper examines the environmental footprint of modern AI systems across energy consumption, carbon emissions, water usage, and electronic waste over the full lifecycle of large language models, multimodal foundation models, and agentic workflows, from hardware fabrication and training through fine-tuning and inference to end-of-life disposal. This work provides a conceptual analysis, utilizing order-of-magnitude estimations based on published data, without conducting original physical measurements. We contribute a lifecycle taxonomy that crosses lifecycle phases with five impact dimensions and emission scopes; a Sustainability Assessment Framework for AI Systems (SAFIA) comprising nine indicators; a comparative analysis of traditional, generative, and agentic AI; seven open challenges; policy recommendations for regulators, cloud providers, hardware manufacturers, and AI developers; and a research roadmap to 2035. The analysis indicates that inference can rival or exceed training energy over a deployment lifetime, that agentic workflows can multiply the energy of equivalent single-pass inference by one to several orders of magnitude depending on the number of model and tool calls, that indirect water use and embodied carbon are systematically underreported, and that measurement tools, disclosure practices, and regulation have not kept pace with agentic deployment. These findings call for agent-aware energy measurement, lifecycle-based carbon and water accounting, and mandatory disclosure for large-scale AI training and deployment.
\end{abstract}


\section{Introduction}
\label{sec:introduction}

\begin{table}[htbp!]
\scriptsize
\centering
\caption{List of Acronyms and Abbreviations}
\label{tab:acronyms}
\begin{tabular}{ll}
\hline
\textbf{Abbreviation} & \textbf{Meaning} \\
\hline
AI & Artificial Intelligence \\
API & Application Programming Interface \\
AWQ & Activation-Aware Weight Quantization \\
CAI & Constitutional AI \\
CO$_2$e & Carbon Dioxide Equivalent \\
CPU & Central Processing Unit \\
CRAC & Computer Room Air Conditioner \\
CRAH & Computer Room Air Handler \\
DPO & Direct Preference Optimization \\
DRC & Democratic Republic of the Congo \\
EC & Exclusion Criterion \\
EU & European Union \\
FLOP & Floating-Point Operation \\
GHG & Greenhouse Gas \\
GPU & Graphics Processing Unit \\
GRI & Global Reporting Initiative \\
HFC & Hydrofluorocarbon \\
HVAC & Heating, Ventilation, and Air Conditioning \\
IC & Inclusion Criterion \\
ICT & Information and Communication Technology \\
IEC & International Electrotechnical Commission \\
IoT & Internet of Things \\
ISO & International Organization for Standardization \\
IT & Information Technology \\
JEDEC & Joint Electron Device Engineering Council \\
LCA & Life Cycle Assessment \\
LLM & Large Language Model \\
LoRA & Low-Rank Adaptation \\
ML & Machine Learning \\
MoE & Mixture of Experts \\
NIST & National Institute of Standards and Technology \\
NLP & Natural Language Processing \\
NVML & NVIDIA Management Library \\
OC & Open Challenge \\
PCB & Printed Circuit Board \\
PEFT & Parameter-Efficient Fine-Tuning \\
PFC & Perfluorocarbon \\
PPA & Power Purchase Agreement \\
PPO & Proximal Policy Optimization \\
PUE & Power Usage Effectiveness \\
RAG & Retrieval-Augmented Generation \\
RAPL & Running Average Power Limit \\
RLHF & Reinforcement Learning from Human Feedback \\
RQ & Research Question \\
SAFIA & Sustainability Assessment Framework for AI Systems \\
SFT & Supervised Fine-Tuning \\
TPU & Tensor Processing Unit \\
WUE & Water Usage Effectiveness \\
\hline
\end{tabular}
\end{table}

\subsection{From Generative to Agentic AI}
\label{subsec:genai_agentic}

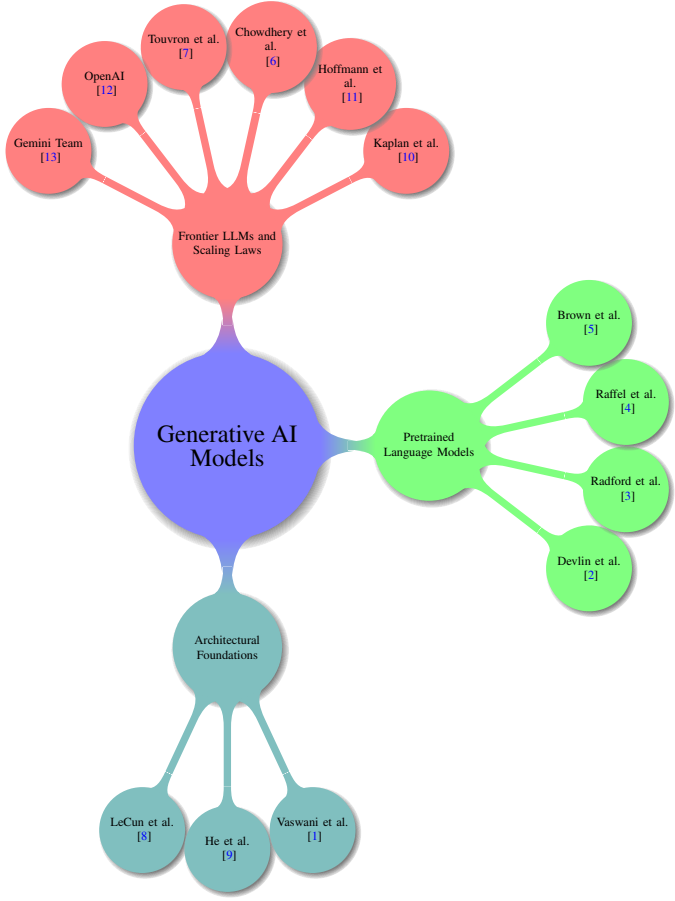
\begin{figure}[htbp]
\centering
\resizebox{\columnwidth}{!}{%
\begin{tikzpicture}[
  mindmap,
  every node/.style={concept, circular drop shadow, minimum size=0.1cm},
  grow cyclic,
  align=flush center,
  concept color=blue!50,
  level 1/.append style={
    sibling angle=90,
    level distance=4cm,
    font=\scriptsize
  },
  level 2/.append style={
    sibling angle=25,
    level distance=4cm,
    font=\scriptsize
  }
]
\node[concept color=blue!50] {\Large Generative AI Models}
  child[concept color=teal!50] { node {Architectural Foundations}
    child { node {LeCun et al.\\\cite{lecun2015deep}} }
    child { node {He et al.\\\cite{he2016deep}} }
    child { node {Vaswani et al.\\\cite{vaswani2017attention}} }
  }
  child[concept color=green!50] { node {Pretrained Language Models}
    child { node {Devlin et al.\\\cite{devlin2019bert}} }
    child { node {Radford et al.\\\cite{radford2019language}} }
    child { node {Raffel et al.\\\cite{raffel2020exploring}} }
    child { node {Brown et al.\\\cite{brown2020language}} }
  }
  child[concept color=red!50] { node {Frontier LLMs and Scaling Laws}
    child { node {Kaplan et al.\\\cite{kaplan2020scaling}} }
    child { node {Hoffmann et al.\\\cite{hoffmann2022training}} }
    child { node {Chowdhery et al.\\\cite{chowdhery2023palm}} }
    child { node {Touvron et al.\\\cite{touvron2023llama}} }
    child { node {OpenAI\\\cite{openai2023gpt4}} }
    child { node {Gemini Team\\\cite{team2023gemini}} }
  };
\end{tikzpicture}
}
\caption{Classification of Generative AI Architectures, Model Families, and Scaling Studies by Category}
\label{fig:mm_models}
\end{figure}

\IEEEPARstart{T}{he} emergence of large-scale generative AI represents a major technological transition. Beginning with the Transformer architecture~\cite{vaswani2017attention} in 2017 and accelerating with BERT~\cite{devlin2019bert}, GPT-2~\cite{radford2019language}, T5~\cite{raffel2020exploring}, GPT-3~\cite{brown2020language}, PaLM~\cite{chowdhery2023palm}, LLaMA~\cite{touvron2023llama}, and subsequent frontier models, generative AI systems have scaled from tens of millions to hundreds of billions of parameters within less than a decade. This rapid scaling, while yielding remarkable capabilities in language understanding, code generation, image synthesis, and scientific reasoning, has been accompanied by escalating computational and infrastructural demands. Figure~\ref{fig:mm_models} organizes the architectures, model families, and scaling studies that frame this evolution.

A qualitatively distinct and more recent development is the emergence of agentic AI: autonomous systems that orchestrate multiple AI models, tools, and computational processes to achieve complex, multi-step objectives with minimal human intervention~\cite{wang2024survey}. Unlike single-pass inference systems, multi-agent frameworks such as AutoGen~\cite{wu2023autogen} operate through iterative reasoning loops, tool invocations, web browsing, code execution, memory retrieval, and inter-agent communication cycles that can span a large number of LLM calls per task~\cite{shen2023hugginggpt}. Because agentic deployments are becoming increasingly prevalent in enterprise, scientific, and consumer applications, their resource demands have become a pressing environmental question.

\subsection{Environmental Challenges and Knowledge Gaps}
\label{subsec:env_gaps}

The environmental challenges posed by these systems are multifaceted and interconnected. Energy consumption is the most widely studied dimension, and the emissions footprint of ICT, including data centers, has been projected to grow substantially~\cite{belkhir2018assessing}. Training a single LLM can consume hundreds of megawatt-hours of electricity or more and emit hundreds of tonnes of CO$_2$e, depending on the energy mix of the host data center, and Luccioni et al.~\cite{luccioni2022estimating} showed that such estimates are highly sensitive to the carbon intensity of the local electricity grid. Inference-phase demands, given the scale of deployment, may rival or exceed training costs over a model's operational lifetime. Carbon emissions associated with AI computation are a direct function of energy consumption and grid carbon intensity. The geographic distribution of data center operations is therefore a first-order environmental variable: the same workload can emit more than an order of magnitude more CO$_2$e on a coal-dominated grid than on a grid relying on hydroelectric or nuclear power. 

The footprint of AI also extends beyond electricity. The fabrication of GPUs, TPUs, and emerging AI accelerators requires rare earth elements, water, and energy-intensive chemical processes. Water consumption for data center cooling has received less attention than energy, yet it represents a substantial and growing burden, and reported water metrics typically exclude the indirect water embedded in electricity generation and semiconductor fabrication. These burdens are compounded by the rise of agentic AI, whose environmental implications remain largely uncharacterized: existing measurement and reporting tools focus on single-model training and inference and do not address the compounding energy overhead introduced by tool-use loops, RAG pipelines~\cite{lewis2020retrieval}, chain-of-thought reasoning~\cite{wei2022chain}, and multi-agent coordination. The obsolescence of AI hardware, driven by frequent refresh, further exacerbates e-waste generation, a globally significant environmental hazard.

Several studies have addressed aspects of AI's environmental impact. Schwartz et al.~\cite{schwartz2020greenai} introduced the concept of ``Green AI'' and called for reporting computational costs alongside accuracy metrics, Anthony et al.~\cite{anthony2020carbontracker} developed the CarbonTracker tool for real-time energy monitoring during training, and Wu et al.~\cite{wu2022sustainable} offered a broad, hardware-oriented analysis of the environmental implications of AI. However, these works share several limitations. First, none provides a systematic analysis that jointly covers energy, carbon, water, and e-waste within a unified framework. Second, none addresses agentic AI architectures, which introduce qualitatively different computational patterns. Third, the evolving landscape of model architectures, hardware platforms, and deployment patterns means that even reviews published in 2022--2023 are partially obsolete with respect to 2024--2025 model families. Fourth, existing reviews rarely engage with LCA methodology, leaving embodied carbon and hardware-phase impacts analytically separated from operational impacts~\cite{freitag2021real}.

\subsection{Contributions and Organization}
\label{subsec:contributions}

To address these gaps, this review makes the following contributions:

\begin{enumerate}[leftmargin=*, label=\textbf{C\arabic*.}]
\item \textbf{Lifecycle Impact Taxonomy:} We develop an environmental impact taxonomy for generative and agentic AI that crosses lifecycle phases with five impact dimensions.

\item \textbf{Agentic AI Environmental Analysis:} We analyze the environmental footprint of agentic AI systems, including multi-agent workflows, tool-use loops, and autonomous reasoning chains, and formalize their overhead through an agentic energy multiplier.

\item \textbf{SAFIA Framework:} We propose SAFIA, a nine-indicator assessment framework applicable to both training and deployment phases.

\item \textbf{Comparative Environmental Analysis:} We compare traditional AI, generative AI, and agentic AI along a common set of environmental dimensions, highlighting large differences in per-task environmental burden driven chiefly by the agentic multiplier.

\item \textbf{Open Challenges and Roadmap:} We identify seven open challenges and organize them into a research roadmap extending to 2035.

\item \textbf{Policy Framework:} We develop policy recommendations for regulators, cloud providers, hardware manufacturers, and AI developers.
\end{enumerate}

The remainder of this paper is organized as follows. Section~\ref{sec:methodology} presents the research methodology and PRISMA workflow, and Section~\ref{sec:taxonomy} introduces the environmental impact taxonomy. Sections~\ref{sec:energy}, \ref{sec:carbon}, and~\ref{sec:water_ewaste} analyze energy, carbon, and the water and e-waste footprints, respectively. Section~\ref{sec:comparison} compares traditional, generative, and agentic AI, Section~\ref{sec:green_ai} reviews mitigation strategies, and Section~\ref{sec:framework} presents the SAFIA framework. Section~\ref{sec:challenges_policy} discusses open challenges, the research roadmap, policy implications, and the limitations of this review, and Section~\ref{sec:conclusion} concludes. Table~\ref{tab:acronyms} lists the acronyms used throughout the paper.

\section{Research Methodology}
\label{sec:methodology}

This review follows an explicit protocol defined by its research questions, search strategy, and selection criteria, which together determine the body of evidence analyzed in the remainder of the paper.

\subsection{Research Questions}
\label{subsec:rqs}

\begin{description}
\item[\textbf{RQ1}] What are the quantified energy consumption profiles of generative AI and agentic AI systems across training, fine-tuning, inference, and agentic workflow phases?

\item[\textbf{RQ2}] What carbon accounting methodologies exist for AI systems, and how do carbon emissions estimates vary across model families, hardware platforms, and geographic deployments?

\item[\textbf{RQ3}] What is the water footprint of AI systems, and how does it compare across lifecycle phases including hardware fabrication, data center cooling, and electricity generation?

\item[\textbf{RQ4}] What e-waste and material resource depletion impacts are attributable to the AI hardware lifecycle, including manufacturing, use, and end-of-life phases?

\item[\textbf{RQ5}] What mitigation strategies, assessment frameworks, and policy instruments exist or can be developed to reduce the environmental impact of AI systems?
\end{description}

\subsection{Search Strategy and Study Selection}
\label{subsec:search}

To answer these questions, a systematic literature search was conducted across seven electronic databases: Web of Science, Scopus, IEEE Xplore, ACM Digital Library, ScienceDirect, SpringerLink, and arXiv. The search was conducted in 2026 and covered publications from 2015 to 2026. The primary search string was constructed using Boolean operators combining terms from four conceptual clusters:

\textit{Cluster A (AI Systems):} ``large language model'' OR ``generative AI'' OR ``foundation model'' OR ``agentic AI'' OR ``autonomous agent'' OR ``multi-agent system'' OR ``deep learning'' OR ``neural network''

\textit{Cluster B (Environmental Dimensions):} ``energy consumption'' OR ``carbon footprint'' OR ``CO2 emissions'' OR ``water usage'' OR ``electronic waste'' OR ``e-waste'' OR ``lifecycle assessment''

\textit{Cluster C (Computing Infrastructure):} ``data center'' OR ``GPU'' OR ``TPU'' OR ``AI accelerator'' OR ``cloud computing'' OR ``high-performance computing''

\textit{Cluster D (Sustainability):} ``green AI'' OR ``sustainable computing'' OR ``energy efficiency'' OR ``carbon neutrality'' OR ``net zero''

The full search string applied Cluster A AND (Cluster B OR Cluster C) AND/OR Cluster D, with field restrictions to title, abstract, and keywords. Reference lists of included papers were manually screened to identify additional relevant sources. The retrieved records were then screened against the inclusion and exclusion criteria listed in Table~\ref{tab:inclusion_exclusion}.

\begin{table}[!t]
\caption{Inclusion and Exclusion Criteria}
\label{tab:inclusion_exclusion}
\centering
\footnotesize
\setlength{\tabcolsep}{3pt}
\renewcommand{\arraystretch}{1.08}
\begin{tabularx}{\columnwidth}{@{}>{\bfseries}lX@{}}
\toprule
Category & Criterion \\
\midrule
\multicolumn{2}{@{}l}{\textit{Inclusion criteria}} \\
IC1 & Peer-reviewed journal articles, preprints, conference proceedings, or technical reports published 2015--2026 \\
IC2 & Direct measurement, estimation, or analysis of energy consumption, carbon emissions, water usage, or e-waste for AI or ML systems \\
IC3 & Studies addressing AI hardware lifecycle, data center infrastructure, or AI-specific computing infrastructure \\
IC4 & Review, survey, or meta-analysis papers on green AI, sustainable computing, or AI environmental impact \\
IC5 & Policy, regulatory, or governance analyses specifically addressing AI environmental sustainability \\
IC6 & Written in English \\
\midrule
\multicolumn{2}{@{}l}{\textit{Exclusion criteria}} \\
EC1 & Studies addressing only traditional, non-neural AI systems without comparison to modern ML \\
EC2 & Energy efficiency studies focused exclusively on edge devices, mobile phones, or IoT sensors without data center relevance \\
EC3 & Papers without quantitative metrics or measurable environmental indicators \\
EC4 & Duplicate publications \\
EC5 & Opinion pieces, editorials, or perspective papers without empirical or systematic content \\
EC6 & Preprints not subsequently verified as published or cited \\
\bottomrule
\end{tabularx}
\end{table}

\begin{figure}[!t]
\centering
\begin{tikzpicture}[fig]
  \node[plainbox, text width=5.2cm, anchor=north] (db) at (3.21,0) {%
    Records identified via database search\\ \mbox{$n = 4{,}318$}\\[2pt]
    {\scriptsize\setlength{\tabcolsep}{2pt}\renewcommand{\arraystretch}{1}%
    \begin{tabular}{@{}lr@{\hspace{9pt}}lr@{}}
    Web of Science & 1,124 & ScienceDirect & 489\\
    Scopus & 987 & SpringerLink & 312\\
    IEEE Xplore & 734 & arXiv & 160\\
    ACM Digital Library & 512 & & \\
    \end{tabular}}};
  \node[plainbox, text width=2.2cm, anchor=north] (add) at (7.39,0)
    {Additional records via reference screening\\ \mbox{$n = 89$}};
  \node[plainbox, text width=5.2cm, minimum height=1.3cm, below=7mm of db] (scr)
    {Records after duplicate removal: \mbox{$n = 3{,}691$}\\ Records screened (title and abstract): \mbox{$n = 3{,}691$}};
  \node[exclbox, text width=2.2cm] (ex1) at (add.center |- scr.center)
    {Records excluded as irrelevant\\ \mbox{$n = 2{,}944$}};
  \node[plainbox, text width=5.2cm, minimum height=1.6cm, below=7mm of scr] (elig)
    {Full-text articles assessed for eligibility\\ \mbox{$n = 747$}};
  \node[exclbox, text width=2.2cm] (ex2) at (add.center |- elig.center)
    {Full-text articles excluded: \mbox{$n = 658$}\\[1pt]
     {\scriptsize EC1: 112; EC2: 98\\ EC3: 145; EC4: 117\\ EC5: 61; EC6: 125}};
  \node[plainbox, text width=5.2cm, minimum height=1.2cm, below=7mm of elig, font=\footnotesize\bfseries] (inc)
    {Studies included in the synthesis\\ \mbox{$n = 75$}};
  \draw[arr] (db) -- (scr);
  \draw[arr] (scr) -- (elig);
  \draw[arr] (elig) -- (inc);
  \draw[arr] (scr.east) -- (ex1.west);
  \draw[arr] (elig.east) -- (ex2.west);
  \coordinate (m) at ($(db.south)!0.5!(scr.north)$);
  \draw[link] (add.south) |- (m);
  \fill[cCarbon] (m) circle (1.2pt);
  \begin{scope}[on background layer]
    \foreach \fitlist/\lab/\tint in {{(db)(add)}/Identification/8, {(scr)(ex1)}/Screening/13, {(elig)(ex2)}/Eligibility/18, {(inc)}/Included/24} {
      \node[fit=\fitlist, inner sep=0pt] (row) {};
      \path let \p1=(row.north), \p2=(row.south) in
        node[draw=cCarbon!75, fill=cCarbon!\tint, rounded corners=2pt, line width=0.6pt,
             minimum width=0.40cm, minimum height={\y1-\y2}, inner sep=0pt, anchor=north west] at (0,\y1) {}
        node[rotate=90, font=\scriptsize\bfseries] at (0.20,{0.5*(\y1+\y2)}) {\lab};
    }
  \end{scope}
\end{tikzpicture}
\caption{PRISMA flow of the literature search.}
\label{fig:prisma}
\end{figure}
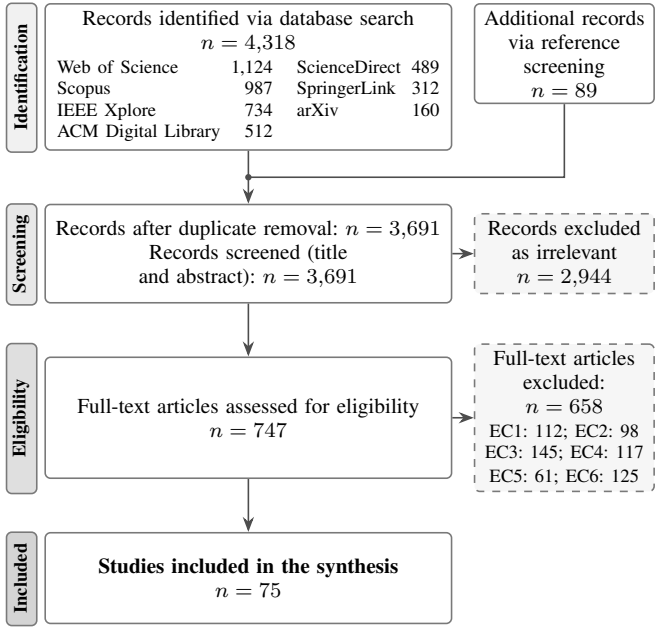

Study selection followed the PRISMA 2020 framework~\cite{page2021prisma} to ensure methodological rigor and transparency. Figure~\ref{fig:prisma} shows the complete workflow, from identification and duplicate removal through title and abstract screening and full-text eligibility assessment to the studies included in the synthesis. The final synthesis comprises 75 studies, and Figure~\ref{fig:prisma} reports the number of records excluded at each stage together with the criterion-level reasons for full-text exclusion.

To separate evidence from interpretation, findings reported in the literature are attributed to their sources through citations, whereas the taxonomy, the agentic multiplier, SAFIA, and the qualitative levels of the comparative analysis constitute the authors' synthesis; forward-looking statements are framed as open challenges, and policy recommendations are presented as proposals. Energy, carbon, and water values are estimates published by the cited studies, derived from hardware power, runtime, and efficiency factors; no measurements were made for this review, and the SAFIA weights and disclosure thresholds are our own proposals.

\section{Taxonomy of Environmental Impacts}
\label{sec:taxonomy}

The studies retained through this protocol cover energy, carbon, water, and materials, but they rarely connect these dimensions. A core contribution of this review is therefore a unified environmental impact taxonomy that organizes this fragmented literature into a coherent analytical structure, which also frames the analysis in the rest of the paper.

\begin{figure}[htbp]
\centering
\resizebox{\columnwidth}{!}{%
\begin{tikzpicture}[
  mindmap,
  every node/.style={concept, circular drop shadow, minimum size=0.1cm},
  grow cyclic,
  align=flush center,
  concept color=blue!50,
  level 1/.append style={
    sibling angle=90,
    level distance=4cm,
    font=\scriptsize
  },
  level 2/.append style={
    sibling angle=25,
    level distance=4cm,
    font=\scriptsize
  }
]
\node[concept color=blue!50] {\Large Environmental Footprint of AI}
  child[concept color=teal!50] { node {Energy and Carbon Accounting}
    child { node {Strubell et al.\\\cite{strubell2019energy}} }
    child { node {Schwartz et al.\\\cite{schwartz2020greenai}} }
    child { node {Henderson et al.\\\cite{henderson2020towards}} }
    child { node {Lottick et al.\\\cite{lottick2019energy}} }
    child { node {Bannour et al.\\\cite{bannour2021evaluating}} }
    child { node {Dodge et al.\\\cite{dodge2022measuring}} }
    child { node {Lannelongue et al.\\\cite{lannelongue2021green}} }
  }
  child[concept color=green!50] { node {Deployment and Infrastructure}
    child { node {Luccioni et al.\\\cite{luccioni2022bloom}} }
    child { node {Jouppi et al.\\\cite{jouppi2017datacenter}} }
    child { node {Masanet et al.\\\cite{masanet2020recalibrating}} }
    child { node {Belkhir and Elmeligi\\\cite{belkhir2018assessing}} }
  }
  child[concept color=red!50] { node {Water, Materials, and Lifecycle}
    child { node {Li et al.\\\cite{li2023making}} }
    child { node {Grubert and Sanders\\\cite{grubert2020water}} }
    child { node {Gupta et al.\\\cite{gupta2022act}} }
    child { node {Forti et al.\\\cite{forti2020global}} }
    child { node {Bald\'e et al.\\\cite{baldé2017global}} }
    child { node {Crawford\\\cite{crawford2021atlas}} }
    child { node {Freitag et al.\\\cite{freitag2021real}} }
  };
\end{tikzpicture}
}
\caption{Classification of Environmental Footprint Studies on AI by Focus Area}
\label{fig:mm_evidence}
\end{figure}
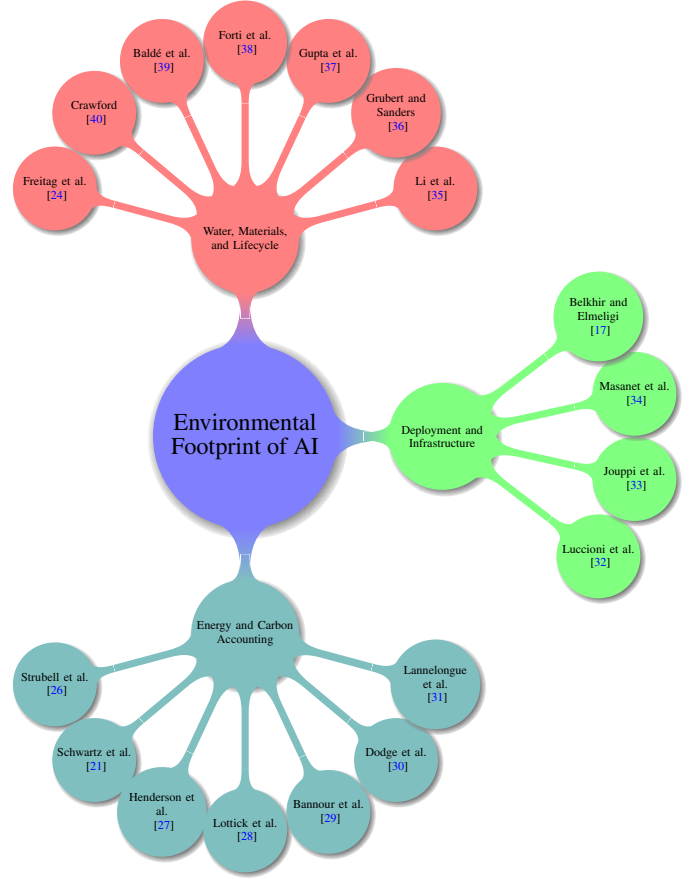

Existing characterizations tend to treat energy, carbon, water, and e-waste as separate analytical silos, despite their physical interconnections: the electricity consumed by data centers simultaneously drives carbon emissions through grid generation and water consumption through cooling and thermoelectric generation, while the hardware that consumes it carries embodied carbon and eventually becomes e-waste. Figure~\ref{fig:mm_evidence} illustrates this fragmentation by classifying the environmental footprint studies reviewed here according to their focus.

The taxonomy crosses seven lifecycle phases (hardware manufacturing, infrastructure build, which covers data center construction and system assembly, model training, fine-tuning, inference, agentic workflow, and hardware decommissioning) with five impact dimensions, and annotates each phase with its dominant GHG scope. Figure~\ref{fig:taxonomy} presents this structure.

\begin{figure}[!t]
\centering
\begin{tikzpicture}[fig]
  \def\rh{0.50}   
  \def\cw{0.97}   
  \def\xa{4.24}   
  \def\xs{0.24}   
  \def\yo{0.10}   
  \begin{scope}[on background layer]
    \fill[cAgent!12, draw=cAgent!75, line width=0.6pt, rounded corners=2pt]
      (-0.02,{-5*\rh-\yo+0.03}) rectangle ({\xa+4*\cw+0.5},{-6*\rh-\yo-0.03});
  \end{scope}
  \node[font=\scriptsize\bfseries, align=center, anchor=south] at (\xs,0.02) {GHG\\scope};
  \node[font=\scriptsize\bfseries, anchor=south west] at (0.52,0.02) {Lifecycle phase};
  \foreach \c/\t [count=\j from 0] in {cEnergy/Energy, cCarbon/Carbon, cWater/Water, cMaterial/{Materials\\and\\e-waste}, cLand/{Land\\use}} {
    \node[box=\c, font=\scriptsize, minimum width=0.93cm, minimum height=0.86cm, inner sep=0.5pt, anchor=south]
      at ({\xa+\j*\cw},0) {\t};
  }
  \draw[cCarbon!75, line width=0.6pt] (-0.02,-0.04) -- ({\xa+4*\cw+0.5},-0.04);
  \foreach \lab/\sc/\va/\vb/\vc/\vd/\ve [count=\i] in {
      Hardware manufacturing/S3/1/1/1/1/0,
      Infrastructure build/S3/0/0/1/0/1,
      Model training/S2/1/1/1/0/0,
      Fine-tuning/S2/0/0/0/0/0,
      Inference/S2/1/1/0/0/0,
      {Agentic workflow ($\times\mu_A$)}/S2/1/1/0/0/0,
      Hardware decommissioning/S3/0/0/0/1/0} {
    \pgfmathsetmacro{\y}{-(\i-0.5)*\rh-\yo}
    \node[font=\scriptsize\bfseries, text height=1.5ex, text depth=0.25ex] at (\xs,\y) {\sc};
    \node[anchor=west, text height=1.6ex, text depth=0.3ex] at (0.52,\y) {\lab};
    \foreach \v/\c [count=\j from 0] in {\va/cEnergy, \vb/cCarbon, \vc/cWater, \vd/cMaterial, \ve/cLand} {
      \ifnum\v=1
        \node[primary=\c] at ({\xa+\j*\cw},\y) {};
      \else
        \node[secondary=\c] at ({\xa+\j*\cw},\y) {};
      \fi
    }
  }
  \draw[cCarbon!75, line width=0.6pt] (-0.02,{-7*\rh-\yo-0.08}) -- ({\xa+4*\cw+0.5},{-7*\rh-\yo-0.08});
  \pgfmathsetmacro{\yl}{-7*\rh-\yo-0.36}
  \node[primary=cCarbon] (lp) at (0.12,\yl) {};
  \node[anchor=west, font=\scriptsize] at (lp.east) {Primary impact};
  \node[secondary=cCarbon] (ls) at (2.30,\yl) {};
  \node[anchor=west, font=\scriptsize] at (ls.east) {Secondary or indirect impact};
  \filldraw[fill=cAgent!12, draw=cAgent!75, line width=0.6pt, rounded corners=1pt] (6.02,{\yl-0.1}) rectangle (6.42,{\yl+0.1});
  \node[anchor=west, font=\scriptsize] at (6.42,\yl) {Agentic phase};
  \node[anchor=west, font=\scriptsize] at (-0.06,{\yl-0.34}) {S1: direct emissions; S2: purchased electricity; S3: value chain};
\end{tikzpicture}
\caption{Lifecycle taxonomy of the environmental impacts of AI systems}
\label{fig:taxonomy}
\end{figure}

This taxonomy reveals several relationships. Hardware manufacturing, largely invisible in operational carbon accounting, is a primary contributor to energy, carbon, water, and materials for advanced AI chips. Inference, often treated as environmentally negligible compared to training, emerges as a primary energy and carbon contributor when integrated across deployment lifetime and usage volume. Agentic workflows inherit all inference-phase impacts but amplify them through multiplicative looping behavior. The analysis that follows examines these relationships dimension by dimension, beginning with energy, from which most of the other impacts derive.

\section{Energy Consumption of Generative and Agentic AI}
\label{sec:energy}

Energy is the dimension from which most other impacts follow, because carbon emissions and much of the water footprint scale with electricity use. This section follows energy through the lifecycle phases of the taxonomy, from training and fine-tuning to inference and agentic workflows, and then synthesizes the key drivers of consumption.

\subsection{Training and Fine-Tuning}
\label{subsec:training}

The training phase of large AI models is the most energy-intensive single computational event in the model lifecycle ~\cite{infollm}, although it is not necessarily the largest cumulative energy consumer over the deployment lifetime. Training energy is governed by the number of model parameters, the number of training tokens, and the efficiency of the accelerator cluster. The scaling laws characterized by Kaplan et al.~\cite{kaplan2020scaling}, in which per-token forward-pass compute grows approximately linearly with parameter count, and later refined by Hoffmann et al.~\cite{hoffmann2022training}, relate model performance to the compute budget through a power law, with model size and training tokens ideally scaled in equal proportion. Training compute, and therefore energy, thus grows approximately linearly with parameter count for a fixed number of training tokens. In practice, several recent models, such as LLaMA, have been trained on more tokens than compute-optimal ratios suggest, which increases training energy.

Early estimates by Strubell et al.~\cite{strubell2019energy} quantified the energy and carbon costs of training NLP models and showed that the CO$_2$e emitted by a full neural architecture search for a Transformer model could be comparable to the lifetime emissions of several average cars. These estimates were subsequently refined and extended to larger models; for GPT-3, trained on V100 GPUs, the published estimates of training energy and emissions are approximately 1,287~MWh and 552~tonnes of CO$_2$e under the average grid mix of the United States. Table~\ref{tab:energy_models} lists training energy and emissions of representative models; the contrast between GPT-3 and BLOOM illustrates how strongly emissions depend on the grid mix of the host data center.

\begin{table}[!t]
\caption{Reported Training Energy and Emissions of Representative Models}
\label{tab:energy_models}
\centering
\footnotesize
\setlength{\tabcolsep}{3.5pt}
\begin{tabular}{@{}lllrr@{}}
\toprule
\textbf{Model} & \textbf{Parameters} & \textbf{Hardware} & \textbf{Energy (MWh)} & \textbf{CO$_2$e (t)} \\
\midrule
T5-11B & 11B & TPU v3 & 86 & 47 \\
GPT-3 & 175B & V100 GPUs & 1,287 & 552\textsuperscript{a} \\
BLOOM & 176B & A100 GPUs & 433 & 24.7\textsuperscript{b} \\
\bottomrule
\multicolumn{5}{@{}p{0.98\columnwidth}@{}}{\scriptsize\itshape Values are published estimates, not measured. Values are grid-dependent and not directly comparable across rows.} \\
\end{tabular}
\end{table}

These estimates rest on heterogeneous methods. Some studies measure hardware power directly, while others derive energy from FLOP counts and hardware specifications or apply estimation tools such as CarbonTracker or ML CO$_2$ Impact. This heterogeneity produces estimates that can differ substantially for the same model, a critical source of scientific uncertainty. A further complication is that frontier model providers have not publicly disclosed the full training compute, energy consumption, or carbon emissions of several recent frontier models~\cite{openai2023gpt4,team2023gemini}. This opacity limits independent scientific assessment and underscores the need for the mandatory energy disclosure requirements that we recommend later in this paper.

Once pretrained, models are usually adapted before deployment. Fine-tuning adapts a pretrained model to specific tasks or domains using a smaller, curated dataset, typically through supervised learning on labeled examples or RLHF~\cite{ziegler2019fine}. Because fine-tuning processes far less data than pretraining and, in PEFT approaches, trains only small adapter modules~\cite{houlsby2019parameter}, its energy requirements are substantially lower than those of full training.

Reported energy costs for fine-tuning are sparse but consistent in direction. In the RLHF pipeline described by Ouyang et al.~\cite{ouyang2022training}, a SFT phase, a reward model training phase, and PPO iterations each require separate forward and backward passes over the base model, yet the total alignment compute remains a small fraction of the pretraining compute. LoRA~\cite{hu2022lora} reduces the number of trainable parameters by more than 99\% while preserving downstream performance, potentially reducing fine-tuning energy substantially. However, the proliferation of fine-tuning at scale, with a large number of task-specific model variants trained by enterprises and individual practitioners, creates an aggregate energy burden that may be non-trivial. CAI~\cite{bai2022constitutional} and DPO~\cite{rafailov2023direct} aim to reduce the complexity of RLHF, potentially with corresponding energy reductions, although empirical energy comparisons between alignment methods remain scarce.

\subsection{Inference}
\label{subsec:inference}

Once a model is deployed, its energy footprint shifts from one-time training to continuous serving. The inference phase, which generates outputs for individual user queries, is often characterized as energetically negligible per query relative to training. This framing is misleading at scale. Patterson et al.\cite{patterson2021carbon} noted that most organizations spend more energy serving models than training them, and Desislavov et al.\cite{desislavov2023trends} showed that, because of the multiplicative factor of deployment, inference accounts for most of the computing effort of widely used models. Single-query energy consumption varies widely across model sizes, tasks, and hardware configurations, as the per-query measurements of Luccioni et al.\cite{luccioni2022bloom} show.

Inference efficiency is further affected by batching strategies, numerical precision, speculative decoding~\cite{leviathan2023fast}, and serving infrastructure choices such as PagedAttention~\cite{kwon2023efficient}. PagedAttention and FlashAttention~\cite{dao2022flashattention} are hardware-aware optimizations that reduce inference memory footprint and latency but do not necessarily reduce per-query energy proportionally. Multimodal inference, which combines text, image, audio, and video, adds further variability.

\subsection{Agentic Workflows}
\label{subsec:agentic_workflows}

\begin{figure}[htbp]
\centering
\resizebox{\columnwidth}{!}{%
\begin{tikzpicture}[
  mindmap,
  every node/.style={concept, circular drop shadow, minimum size=0.1cm},
  grow cyclic,
  align=flush center,
  concept color=blue!50,
  level 1/.append style={
    sibling angle=90,
    level distance=4cm,
    font=\scriptsize
  },
  level 2/.append style={
    sibling angle=25,
    level distance=4cm,
    font=\scriptsize
  }
]
\node[concept color=blue!50] {\Large Agentic AI Workflows}
  child[concept color=teal!50] { node {Reasoning and Retrieval Patterns}
    child { node {Wei et al.\\\cite{wei2022chain}} }
    child { node {Yao et al.\\\cite{yao2023react}} }
    child { node {Shinn et al.\\\cite{shinn2023reflexion}} }
    child { node {Yao et al.\\\cite{yao2023tree}} }
    child { node {Lewis et al.\\\cite{lewis2020retrieval}} }
  }
  child[concept color=green!50] { node {Agent Frameworks and Orchestration}
    child { node {Wu et al.\\\cite{wu2023autogen}} }
    child { node {Shen et al.\\\cite{shen2023hugginggpt}} }
    child { node {Park et al.\\\cite{park2023generative}} }
    child { node {Wang et al.\\\cite{wang2024survey}} }
  }
  child[concept color=red!50] { node {Agent Benchmarks}
    child { node {Jimenez et al.\\\cite{jimenez2023swe}} }
    child { node {Mialon et al.\\\cite{mialon2023gaia}} }
    child { node {Liu et al.\\\cite{liu2023agentbench}} }
  };
\end{tikzpicture}
}
\caption{Classification of Agentic AI Patterns, Frameworks, and Benchmarks by Category}
\label{fig:mm_agentic}
\end{figure}
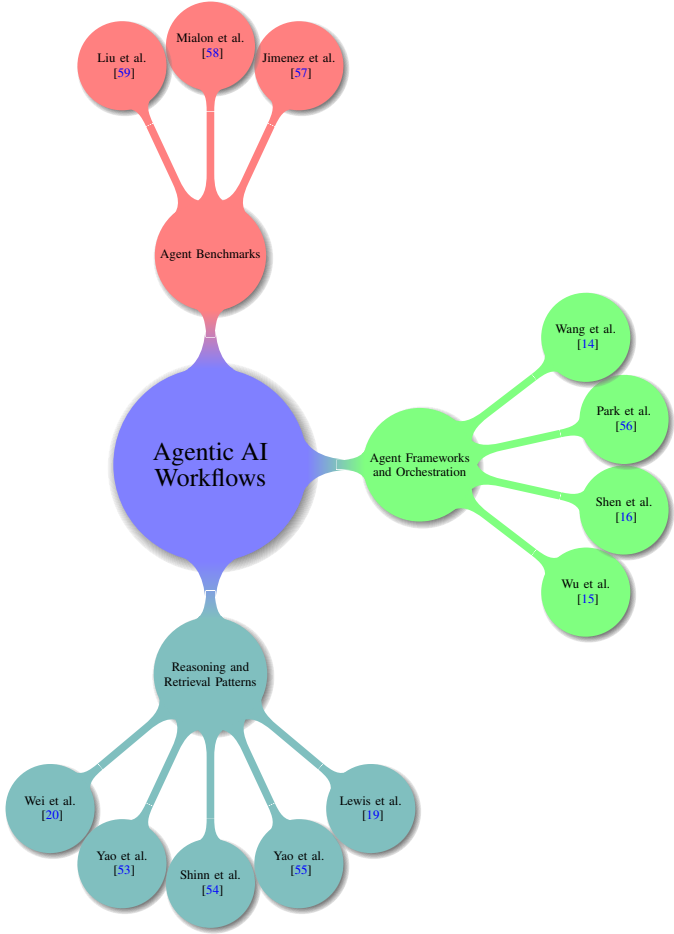

Agentic systems build on inference but change its structure, and with it their environmental implications. Whereas conventional LLM inference answers each request with a single model invocation, agentic systems engage in iterative reasoning-action loops, with ReAct~\cite{yao2023react}, Reflexion~\cite{shinn2023reflexion}, and Tree of Thoughts~\cite{yao2023tree} as archetypal patterns, in which the model calls tools, retrieves information, evaluates intermediate results, and revises its approach over multiple LLM invocations. Figure~\ref{fig:mm_agentic} classifies the reasoning patterns, agent frameworks, and benchmarks that shape these workflows.

The environmental implications of this multiplicative behavior are not captured by the measurement tools and reporting frameworks reviewed here. We characterize the \textit{agentic energy multiplier} $\mu_A$ as:

\begin{equation}
\label{eq:agentic_multiplier}
\begin{split}
E_{\text{agentic}} &= \mu_A \cdot E_{\text{inference}} \\
                   &= \left(N_{\text{calls}} + N_{\text{tools}} \cdot \omega_{\text{tool}}\right) \cdot E_{\text{single}}
\end{split}
\end{equation}

where $N_{\text{calls}}$ is the number of LLM invocations per agentic task, $N_{\text{tools}}$ is the number of tool calls, $\omega_{\text{tool}}$ is the energy cost weight of tool execution, and $E_{\text{single}}$ is the energy cost of a single LLM call. As an indicative, qualitative anchoring, $\omega_{\text{tool}}$ is low for local deterministic calls such as arithmetic or lookup APIs, medium for embedding generation and vector retrieval in RAG pipelines, and high for remote code compilation and execution, graphics rendering, or search over very large corpora; calibrating these levels requires instrumented measurement.

$N_{\text{calls}}$ ranges from a few calls for simple task completion to far more for complex, long-horizon tasks such as resolving real-world software issues in SWE-bench\cite{jimenez2023swe}. Depending on the number of model and tool calls, $\mu_A$ can therefore reach one to several orders of magnitude. When tool execution costs are included, particularly for computationally intensive tools such as code interpreters or search over large corpora, $\mu_A$ may be further amplified.

Multi-agent systems amplify this effect further. In AutoGen-based frameworks, a single task may be decomposed across several specialized agents, each independently executing reasoning loops, with inter-agent communication adding overhead. Park et al.~\cite{park2023generative} demonstrated a 25-agent simulation in which each agent executed multiple daily reasoning cycles; although this was a research demonstration, it illustrates the potential scale of agentic energy consumption in deployed systems. The absence of standardized benchmarks for agentic energy consumption is a critical gap: existing trackers such as CarbonTracker measure GPU energy during training or single-model inference but cannot follow the distributed, asynchronous, multi-model energy profile of an agentic pipeline.

\subsection{Key Drivers of Energy Consumption}
\label{subsec:energy_drivers}

Synthesizing the preceding analysis, we identify six primary drivers of AI energy consumption:

\begin{enumerate}[leftmargin=*]
\item \textit{Model size:} Parameter count is the dominant determinant of per-token compute and hence energy, and compression techniques can substantially reduce effective model size with limited performance loss.

\item \textit{Hardware efficiency:} Specialized accelerators such as the TPU~\cite{jouppi2017datacenter} offer efficiency profiles that vary across workload types, and accelerator energy efficiency has improved across hardware generations.

\item \textit{Data center PUE:} PUE is the ratio of total data center energy to IT equipment energy; average PUE has declined substantially over the past decade, with hyperscale facilities achieving particularly low values\cite{masanet2020recalibrating}.

\item \textit{Grid carbon and renewable mix:} Although this is a carbon-intensity rather than an energy driver, it determines the environmental severity of a given energy expenditure, and co-locating compute with renewable generation is a highly effective carbon mitigation strategy.

\item \textit{Algorithmic efficiency:} Attention optimizations, MoE architectures, and efficient decoding reduce redundant computation.

\item \textit{Agentic loop structure:} The multiplicative effect of iterative reasoning loops is a driver without analogue in traditional AI systems; early stopping, caching of intermediate results, and tool-call minimization are critical levers.
\end{enumerate}

Because the environmental cost of this energy depends on how the electricity is generated and on the hardware that consumes it, we turn next to carbon emissions.

\section{Carbon Emissions Associated with AI Systems}
\label{sec:carbon}

Energy use becomes a climate impact through the carbon intensity of the electricity that supplies it and through the emissions embodied in hardware. This section examines how these emissions are accounted for, why they vary, and what current accounting omits.

\subsection{Carbon Accounting Methodologies}
\label{subsec:carbon_methods}

Carbon accounting for AI systems draws on methodologies developed for corporate GHG reporting. In the scope framework that Gupta et al.~\cite{2023chasing} applied to computing, Scope~1 covers direct emissions from owned or controlled sources, Scope~2 covers indirect emissions from purchased electricity, and Scope~3 covers all other indirect emissions across the value chain. For AI systems, Scope~1 emissions are typically minimal. Scope~2 emissions dominate operational carbon accounting and correspond to data center electricity consumption multiplied by the marginal or average grid emission factor of the relevant supply zone. Scope~3 emissions, including those from hardware manufacturing and the cloud supply chain, are rarely quantified in published AI carbon assessments and represent a major source of underreporting.

Henderson et al.~\cite{henderson2020towards} proposed a framework for the systematic reporting of the energy and carbon footprints of ML experiments, and Bannour et al.\cite{bannour2021evaluating} found that available estimation tools vary significantly in the estimates they produce. Table~\ref{tab:carbon_methods} summarizes the principal carbon accounting methods applied in the AI literature.

\begin{table}[!t]
\caption{Carbon Accounting Methods for AI Systems}
\label{tab:carbon_methods}
\centering
\footnotesize
\setlength{\tabcolsep}{3pt}
\begin{tabularx}{\columnwidth}{@{}>{\raggedright\arraybackslash}p{2.3cm}
                               >{\raggedright\arraybackslash}X@{}}
\toprule
\textbf{Method} & \textbf{Description} \\
\midrule
FLOP-based estimation & Energy derived from FLOP counts and hardware efficiency data \\
Power metering (Carbon\-Tracker) & Real-time GPU and CPU power measured via NVML and RAPL, combined with grid emission factors \\
ML CO$_2$ Impact & Web calculator using region and cloud provider data \\
LCA-based & Operational and embodied carbon over the hardware lifecycle \\
Scope-based accounting & Corporate scope accounting applied to computing systems \\
\bottomrule
\end{tabularx}
\end{table}

\subsection{Operational Variability and Embodied Carbon}
\label{subsec:carbon_variability}

The carbon intensity of electricity generation varies strongly across regions and over time within a single grid. Lacoste et al.\cite{lacoste2019quantifying} tabulated regional emission factors spanning more than an order of magnitude, so that identical computational workloads can differ by more than an order of magnitude in emissions depending on location. Temporal variability adds further complexity: carbon intensity on a given grid varies with time of day, season, and weather as renewable generation fluctuates. Dodge et al.~\cite{dodge2022measuring} showed that shifting training workloads toward lower-carbon periods can reduce effective emissions, with gains that depend on job duration and region.

Operational emissions, however, are only part of the picture. Embodied carbon, that is, the GHG emissions associated with hardware manufacturing, transportation, and end-of-life processing, is systematically excluded from most AI carbon assessments. Gupta et al.\cite{gupta2022act} developed an architectural carbon model for computer systems showing that embodied carbon can represent a substantial and sometimes dominant share of lifecycle carbon, depending on utilization and operational lifetime. For systems operated at high utilization on renewable energy, embodied carbon can exceed operational carbon. Carbon, moreover, is not the only resource burden that AI hardware and its operation impose.

\section{Water Footprint and Electronic Waste}
\label{sec:water_ewaste}

Beyond energy and carbon, AI systems draw on freshwater during operation and on finite materials throughout the hardware lifecycle. Both dimensions receive far less attention than energy and carbon and are examined in turn below.

\subsection{Water Footprint}
\label{subsec:water}

Water consumption in AI systems arises through two primary pathways, direct cooling of data center IT equipment and indirect water embedded in electricity generation, and through a third, less analyzed pathway: water used in semiconductor fabrication. Data centers remove the waste heat of IT equipment through air cooling, evaporative cooling towers, or liquid cooling. Evaporative systems, common in large facilities because of their energy efficiency, consume freshwater directly through evaporation and blowdown. WUE is defined as annual site water usage divided by IT equipment energy, and it varies widely with cooling technology and climate. Li et al.\cite{li2023making} estimated that training GPT-3 in Microsoft's data centers in the United States directly evaporated approximately 700,000 liters of freshwater, a footprint that grows further once off-site water for electricity generation is included, and noted that the water metrics reported by operators generally exclude such off-site water.

Thermoelectric power generation requires water for condenser cooling, adding an indirect water footprint to every kWh of electricity consumed. Grubert and Sanders\cite{grubert2020water} quantified water withdrawal and consumption factors that vary widely across generation technologies and cooling systems, with once-through cooling, historically common for nuclear and coal plants, having the highest withdrawal rates, and with near-zero operational water consumption for wind and solar photovoltaics. When attributed to AI electricity consumption, this indirect footprint can exceed direct cooling water, particularly for data centers supplied by thermoelectric grids. Renewable electricity therefore provides a water co-benefit beyond carbon reduction, although hydroelectric power has highly variable water consumption depending on reservoir evaporation.

The geographic concentration of large AI data centers in regions experiencing water stress is an underexamined dimension of AI's environmental impact. The tension between data center water demands and agricultural and municipal water needs in such regions represents an emerging social-environmental conflict with no clear regulatory framework.

\subsection{Electronic Waste and Material Resource Depletion}
\label{subsec:ewaste}

Whereas water is consumed mainly during operation, the material footprint of AI spans the whole hardware lifecycle, from mineral extraction to end-of-life processing. Electronic waste from AI systems encompasses the disposal and recycling of computing hardware at the end of useful life. The Global E-waste Monitor~\cite{forti2020global} reported that global e-waste generation reached 53.6 million metric tonnes in 2019 and projected 74.7 million metric tonnes by 2030. AI-specific hardware contributes to this stream through two mechanisms: the rapid obsolescence driven by successive GPU and accelerator generations, and the large scale of deployment in hyperscale data centers. Competitive pressure to deploy each new accelerator generation encourages frequent hardware refresh, generating large quantities of retired hardware.

The environmental impact of e-waste arises from both the hazardous materials contained in retired hardware and the resource inefficiency of not recovering valuable materials. GPUs and AI accelerators contain gold, silver, palladium, and tantalum in their PCBs, alongside potentially hazardous materials including lead solder, cadmium, mercury, and hexavalent chromium~\cite{baldé2017global}. Table~\ref{tab:ewaste} summarizes the environmental impacts associated with each phase of the AI hardware lifecycle.

\begin{table}[!t]
\caption{Environmental Impacts of AI Hardware Across Lifecycle Phases}
\label{tab:ewaste}
\centering
\footnotesize
\setlength{\tabcolsep}{3pt}
\begin{tabularx}{\columnwidth}{@{}>{\raggedright\arraybackslash}p{2.12cm}>{\raggedright\arraybackslash}X>{\raggedright\arraybackslash}p{2.25cm}@{}}
\toprule
\textbf{Phase} & \textbf{Environmental impact} & \textbf{Key materials and substances} \\
\midrule
Raw material extraction & Habitat destruction, water pollution, CO$_2$ emissions from mining & Cobalt (DRC), lithium, tantalum, rare earths \\
Semiconductor fabrication & Water consumption, chemical waste, GHG emissions & HF, NF$_3$, PFCs, ultrapure water \\
PCB manufacturing & Toxic chemical use, worker exposure & Lead, brominated flame retardants, nickel \\
System assembly & Low impact if performed in regulated facilities & Solder, adhesives \\
Operational use & Electricity consumption, heat generation, HVAC wear & Cooling fluids, HFC refrigerants \\
Decommissioning and end of life & Data destruction and transport energy; hazardous release if improperly handled; resource recovery if formally recycled & Au, Ag, Pd, Cu, Pb, Cd, Hg \\
\bottomrule
\end{tabularx}
\end{table}

The manufacturing of advanced AI accelerators also relies on materials facing supply chain vulnerabilities and geopolitical concentration. Cobalt, used in lithium-ion batteries for uninterruptible power supplies in data centers, is primarily mined in the DRC, with associated human rights concerns~\cite{crawford2021atlas}. Gallium and germanium, used in compound semiconductors and some accelerator designs, are subject to Chinese export controls as of 2023. Circular economy approaches, including extended producer responsibility programs, modular hardware design enabling component-level repair and upgrade, and improved precious metal recovery from e-waste recycling, are priority interventions. Taken together, the analyses of energy, carbon, water, and materials show that environmental impacts arise at every lifecycle phase, which raises the question of how they differ across AI paradigms.

\section{Comparative Analysis: Traditional AI vs.\ Generative AI vs.\ Agentic AI}
\label{sec:comparison}

Having examined each dimension separately, we now compare how traditional, generative, and agentic AI differ across them, which requires characterizing their architectural and computational distinctions. Traditional AI systems, including classical ML, shallow neural networks, and early deep learning classifiers with fewer than 100M parameters, are characterized by fixed-size models, deterministic inference paths, and relatively constrained computational requirements~\cite{lecun2015deep}. A ResNet-50 image classifier, for example, requires approximately 4~GFLOPs per inference pass~\cite{he2016deep}.

Generative AI systems, characterized by autoregressive token generation, large parameter counts, and emergent few-shot capabilities, alter this cost structure because output length is variable and directly determines the number of forward passes: generating a response requires one sequential forward pass per output token, and the compute of each pass scales with the parameter count of the model. Per-response cost therefore grows with both output length and model size, and an agentic task multiplies it further by $\mu_A$. Batching mitigates this cost at scale but introduces latency trade-offs.

Table~\ref{tab:comparison} provides a structured, qualitative comparison across the three paradigms. A central finding of this comparison is the near-complete absence of empirical environmental data for agentic AI systems in the published literature. The agentic levels in Table~\ref{tab:comparison} are extrapolated from single-model inference measurements and from the structure of agentic workflows described above; direct measurement of agentic energy consumption across realistic task distributions remains an urgent research priority. These differences also indicate where mitigation can be most effective, which the next section examines.

\begin{table}[!t]
\caption{Qualitative Comparison of Traditional, Generative, and Agentic AI}
\label{tab:comparison}
\centering
\footnotesize
\setlength{\tabcolsep}{3pt}
\begin{tabularx}{\columnwidth}{@{}>{\raggedright\arraybackslash}p{1.95cm}>{\raggedright\arraybackslash}X>{\raggedright\arraybackslash}X>{\raggedright\arraybackslash}X@{}}
\toprule
\textbf{Dimension} & \textbf{Traditional AI} & \textbf{Generative AI} & \textbf{Agentic AI} \\
\midrule
Parameter scale & Small to medium & Large & Large, per agent \\
Training energy & Low to moderate & Very high & Very high \\
Inference energy per task & Low & Moderate to high & Very high \\
Agentic multiplier $\mu_A$ & Not applicable & 1 (single pass) & Greater than 1 \\
Carbon per task & Low & Moderate & High to very high \\
Water per task & Low & Low to moderate & Moderate to high \\
Hardware refresh rate & Slow & Moderate & Rapid \\
E-waste intensity & Low & Moderate & High \\
Carbon reporting & Rarely done & Improving, not standard & Almost never done \\
Mitigation tooling & Mature & Emerging & Very limited \\
LCA coverage & Limited & Partial & Almost none \\
\bottomrule
\multicolumn{4}{@{}p{0.98\columnwidth}@{}}{\scriptsize\itshape Levels are qualitative syntheses of the preceding analyses of energy, carbon, water, and e-waste.} \\
\end{tabularx}
\end{table}

\section{Green AI and Sustainable Mitigation Strategies}
\label{sec:green_ai}

The preceding analysis identifies where environmental impacts arise and how they scale; this section reviews strategies to reduce them, organized from the model level to the infrastructure, agent, and system levels.

\begin{table*}[!t]
\caption{Green AI Mitigation Techniques Grouped by Intervention Level}
\label{tab:green_ai_techniques}
\centering
\footnotesize
\setlength{\tabcolsep}{4pt}
\begin{tabularx}{\textwidth}{@{}l l l X@{}}
\toprule
\textbf{Level} & \textbf{Technique} & \textbf{Mechanism} & \textbf{Expected effect} \\
\midrule
\multirow{6}{*}{Model} & Knowledge distillation & Teacher-student training & Smaller, faster student model with limited accuracy loss \\
 & Quantization  & Reduced numerical precision & Lower memory footprint and higher throughput on compatible hardware \\
 & Structured pruning & Removal of attention heads or layers & Fewer operations per inference with modest accuracy loss \\
 & MoE architectures & Sparse expert activation & Larger capacity at similar per-token compute \\
 & PEFT & Adapter-only fine-tuning & Far fewer trainable parameters \\
 & Efficient attention & Memory-efficient or linear attention & Lower attention memory and latency \\
\midrule
\multirow{3}{*}{Infrastructure} & Carbon-aware scheduling & Temporal workload shifting & Lower emissions; gains depend on job duration and region \\
 & PUE optimization & Advanced cooling & Lower cooling overhead \\
 & Liquid cooling & High-density thermal management & Higher rack density; potentially lower water use \\
\midrule
\multirow{2}{*}{Agent} & Semantic caching & Reuse of results for similar queries & Fewer LLM calls on repetitive workloads \\
 & Early stopping & Confidence-based loop exit & Fewer LLM calls per task \\
\midrule
System & Mandatory reporting & Energy and carbon transparency & Enables measurement and accountability \\
\bottomrule
\end{tabularx}
\end{table*}

\subsection{Model-Level Strategies}
\label{subsec:model_strategies}

\begin{figure}[htbp]
\centering
\resizebox{\columnwidth}{!}{%
\begin{tikzpicture}[
  mindmap,
  every node/.style={concept, circular drop shadow, minimum size=0.1cm},
  grow cyclic,
  align=flush center,
  concept color=blue!50,
  level 1/.append style={
    sibling angle=90,
    level distance=4cm,
    font=\scriptsize
  },
  level 2/.append style={
    sibling angle=25,
    level distance=4cm,
    font=\scriptsize
  }
]
\node[concept color=blue!50] {\Large Model-Level Green AI Techniques}
  child[concept color=teal!50] { node {Model Compression}
    child { node {Hinton et al.\\\cite{hinton2015distilling}} }
    child { node {Sanh et al.\\\cite{sanh2019distilbert}} }
    child { node {Han et al.\\\cite{han2015deep}} }
    child { node {Michel et al.\\\cite{michel2019sixteen}} }
    child { node {Frantar et al.\\\cite{frantar2022gptq}} }
    child { node {Dettmers et al.\\\cite{dettmers2022llm}} }
    child { node {Lin et al.\\\cite{lin2024awq}} }
  }
  child[concept color=green!50] { node {Efficient Architectures and Serving}
    child { node {Fedus et al.\\\cite{fedus2022switch}} }
    child { node {Katharopoulos et al.\\\cite{katharopoulos2020transformers}} }
    child { node {Gu and Dao\\\cite{gu2023mamba}} }
    child { node {Dao et al.\\\cite{dao2022flashattention}} }
    child { node {Kwon et al.\\\cite{kwon2023efficient}} }
    child { node {Leviathan et al.\\\cite{leviathan2023fast}} }
  }
  child[concept color=red!50] { node {Adaptation and Alignment}
    child { node {Houlsby et al.\\\cite{houlsby2019parameter}} }
    child { node {Hu et al.\\\cite{hu2022lora}} }
    child { node {Ziegler et al.\\\cite{ziegler2019fine}} }
    child { node {Ouyang et al.\\\cite{ouyang2022training}} }
    child { node {Bai et al.\\\cite{bai2022constitutional}} }
    child { node {Rafailov et al.\\\cite{rafailov2023direct}} }
  };
\end{tikzpicture}
}
\caption{Classification of Model-Level Green AI Techniques by Category}
\label{fig:mm_efficiency}
\end{figure}
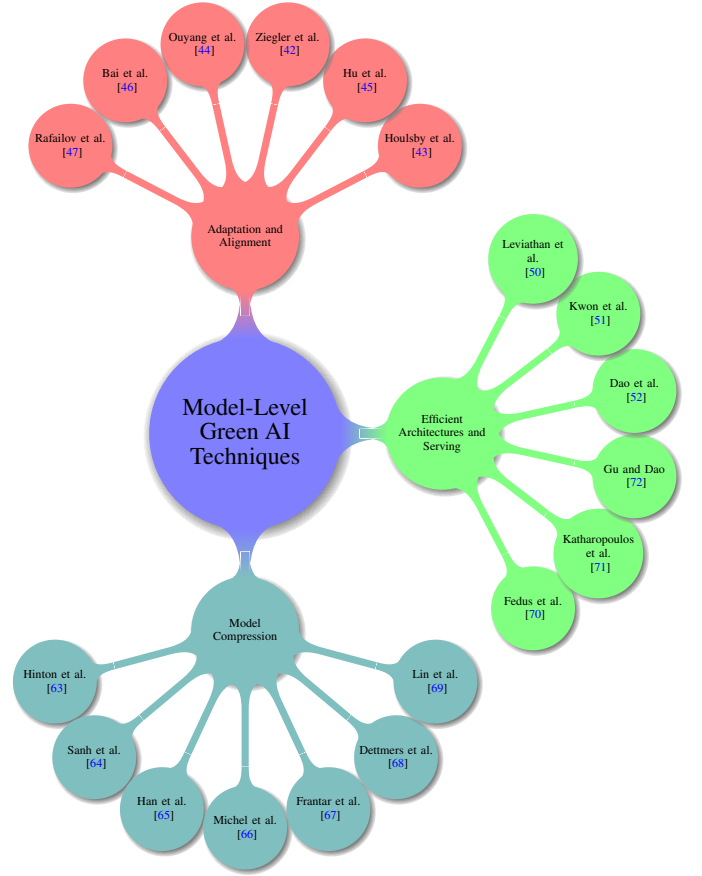

Model-level strategies, which target the computational complexity of models through algorithmic and architectural innovations, form a widely studied category of Green AI interventions\cite{canziani2016analysis}. Figure~\ref{fig:mm_efficiency} classifies the techniques reviewed in this work into model compression, efficient architectures and serving, and adaptation and alignment.

\textit{Knowledge distillation}~\cite{hinton2015distilling}: A large, well-trained teacher model supervises the training of a smaller student model, transferring knowledge at reduced computational cost. Distillation produced DistilBERT\cite{sanh2019distilbert}, a smaller, faster, and cheaper model with 40\% fewer parameters that retains 97\% of the GLUE performance of BERT.

\textit{Quantization}: Reducing the numerical precision of model weights and activations from 32-bit floating point to 16-bit, 8-bit integer, or 4-bit formats reduces memory footprint and can improve inference throughput on compatible hardware. GPTQ~\cite{frantar2022gptq}, LLM.int8()~\cite{dettmers2022llm}, and AWQ~\cite{lin2024awq} are representative post-training quantization methods for LLMs.

\textit{Pruning}: Removing weights~\cite{han2015deep} or attention heads~\cite{michel2019sixteen} from trained models based on magnitude or gradient criteria can substantially reduce effective model size with modest performance degradation. Structured pruning, which removes entire layers or attention heads, is more hardware-friendly than unstructured pruning.

\textit{Efficient architecture design}: MoE architectures~\cite{fedus2022switch} activate only a subset of expert modules per token, achieving large effective capacity with lower per-token FLOPs. Linear attention~\cite{katharopoulos2020transformers} and state-space models such as Mamba~\cite{gu2023mamba} offer sub-quadratic sequence modeling with lower attention compute than standard Transformers.

\textit{Scaling-law alignment}: Training models at compute-optimal budgets, rather than over-computing on model size, can substantially reduce total training energy for equivalent capability.

\subsection{Infrastructure-Level Strategies}
\label{subsec:infra_strategies}

Beyond the model itself, infrastructure-level interventions address the data center and hardware ecosystem in which AI systems operate.

\textit{Renewable energy procurement}: Co-locating or directly powering data centers with renewable energy is a highly effective operational carbon mitigation strategy. ~\cite{bahi2026green}Temporal matching of compute workloads to periods of high renewable availability provides additional benefits without requiring physical co-location ~\cite{BAHIfree}.

\textit{Data center efficiency}: Improving PUE through advanced cooling technologies directly reduces the energy overhead of cooling relative to compute.

\textit{Liquid cooling adoption}: High-density GPU deployments increasingly require liquid cooling because of the thermal limits of air cooling. Direct liquid cooling enables higher rack densities with potentially lower total cooling energy and reduced water consumption compared with evaporative cooling towers.

\textit{Hardware efficiency}: Newer accelerator generations offer higher performance per watt, providing an energy dividend for operators who upgrade hardware, and domain-specific architectures for particular AI workloads, such as inference accelerators and sparse compute engines, offer further efficiency gains.

\subsection{Agent-Level Strategies}
\label{subsec:agent_strategies}

Agent-level sustainability strategies represent a largely unexplored domain, as agent design has not historically considered environmental cost as an optimization objective. We identify the following categories:

\textit{Prompt efficiency}: Longer prompts consume more prefill compute. Compressing context through summarization or selective retrieval reduces per-call energy.

\textit{Caching and memoization}: Caching intermediate reasoning results, retrieved documents, and tool outputs eliminates redundant computation across similar agent tasks. Semantic caching, in which semantically equivalent queries are served from cache without invoking the LLM~\cite{bang2023gptcache}, can substantially reduce the number of LLM calls for repetitive workloads.

\textit{Early stopping and task decomposition}: Confidence-based early stopping in iterative reasoning loops, which stops further LLM calls once the agent has sufficient information, and task decomposition that avoids unnecessary reasoning steps can reduce $N_{\text{calls}}$ with limited effect on task performance in structured settings.

\textit{Footprint-aware agent benchmarking}: Incorporating energy metrics alongside task performance in agentic benchmarks such as SWE-bench, GAIA~\cite{mialon2023gaia}, and AgentBench~\cite{liu2023agentbench} would create incentives for energy-efficient agent design.

\subsection{System-Level Strategies}
\label{subsec:system_strategies}

Technical measures at these three levels operate within a broader ecosystem of policies, incentives, standards, and organizational practices, which system-level strategies address.

\textit{Standardized environmental reporting}: Mandatory reporting of energy consumption, carbon emissions, water usage, and hardware lifecycle data for AI training runs and deployed systems, analogous to financial reporting requirements, would enable investors, regulators, and the public to assess environmental performance~\cite{lottick2019energy}.

\textit{Carbon budgeting for AI projects}: Establishing organizational carbon budgets for AI research and development programs, analogous to financial budgets, incentivizes energy-efficient model and experiment design.

\textit{Open benchmarking}: Public leaderboards that rank AI systems on efficiency metrics alongside accuracy would create market incentives for efficient systems.

\textit{Lifecycle extension programs}: Extending the useful life of AI hardware through firmware updates, secondary market deployment, and improved recycling infrastructure reduces e-waste generation rates.

Table~\ref{tab:green_ai_techniques} summarizes the principal Green AI mitigation techniques, grouped by intervention level. Because these strategies can only be compared and prioritized if their effects are assessed consistently, we next propose an assessment framework.

\section{Proposed SAFIA Framework}
\label{sec:framework}

SAFIA provides this common basis for assessment. It is designed to address the principal limitations of existing approaches: siloed treatment of environmental dimensions, focus on training at the expense of inference and agentic phases, and absence of a lifecycle perspective. Unlike CarbonTracker, ML CO$_2$ Impact, or Green Algorithms, which estimate the energy or carbon of individual computations, SAFIA combines energy, carbon, water, and hardware lifecycle indicators and accounts for the agentic multiplier.

\subsection{Design Principles and Indicators}
\label{subsec:safia_indicators}

SAFIA is organized around three design principles: (1) \textit{completeness}, covering four of the five taxonomy dimensions (energy, carbon, water, and e-waste) across the full lifecycle; (2) \textit{comparability}, producing normalized scores that enable cross-model and cross-deployment comparison; and (3) \textit{practicability}, relying on observable or computable inputs rather than undisclosed proprietary data. Following these principles, SAFIA comprises nine quantifiable indicators organized across the four environmental dimensions. Per-token indicators use 1,000 output tokens as the functional unit, and agentic workloads may additionally be reported per task; lifecycle indicators use one training run or one deployment year. Table~\ref{tab:safia_indicators} defines the nine indicators with their measurement methods and default weights.

\begin{table*}[!t]
\caption{SAFIA Sustainability Assessment Indicators for AI Systems}
\label{tab:safia_indicators}
\centering
\footnotesize
\begin{tabularx}{\textwidth}{@{}l l X l l c@{}}
\toprule
\textbf{Indicator} & \textbf{Dimension} & \textbf{Name} & \textbf{Unit} & \textbf{Measurement method} & \textbf{Default weight} \\
\midrule
I-1 & Energy & Operational energy intensity & Wh per 1,000 output tokens & GPU power metering (NVML, RAPL) & 0.12 \\
I-2 & Energy & Training energy budget & MWh per model version & Hardware logs and PUE factor & 0.12 \\
I-3 & Energy & Agentic energy multiplier $\mu_A$ & Dimensionless & Instrumented agent benchmark & 0.12 \\
I-4 & Carbon & Operational carbon intensity & gCO$_2$e per 1,000 output tokens & I-1 $\times$ grid emission factor & 0.12 \\
I-5 & Carbon & Lifecycle carbon ratio & Embodied/operational fraction & LCA and operational carbon accounting & 0.12 \\
I-6 & Carbon & Renewable energy fraction & \% of operational electricity & Energy attribute certificates & 0.12 \\
I-7 & Water & Direct WUE & L/kWh of IT equipment energy & Data center water meter and IT load & 0.10 \\
I-8 & Water & Indirect water intensity & L per 1,000 output tokens & I-1 $\times$ grid water factor & 0.10 \\
I-9 & E-waste & Hardware lifecycle score & Years (effective) & Mean time to replacement and utilization & 0.08 \\
\bottomrule
\multicolumn{6}{@{}p{0.98\textwidth}@{}}{\scriptsize\itshape Functional units: per-token indicators use 1,000 output tokens; lifecycle indicators use one training run or one deployment year. Default weights sum to 1.00 (energy 0.36, carbon 0.36, water 0.20, e-waste 0.08).} \\
\end{tabularx}
\end{table*}

\subsection{Scoring and Application}
\label{subsec:safia_scoring}

\begin{figure}[!t]
\centering
\begin{tikzpicture}[fig]
  \node[plainbox, text width=8.1cm] (in) at (4.25,0)
    {\textbf{Inputs:} GPU power metering (NVML, RAPL), grid emission and water factors, LCA data, instrumented agent benchmarks};
  \node[box=cEnergy, text width=3.85cm, align=left, anchor=north west] (en)
    at ([yshift=-6.5mm]in.south west)
    {\textbf{Energy} \hfill \mbox{$\sum w = 0.36$}\\[1pt]
     {\scriptsize\begin{tabular*}{3.85cm}{@{}l@{\hspace{3pt}}l@{\extracolsep{\fill}}r@{}}
     I-1 & Operational energy intensity & 0.12\\
     I-2 & Training energy budget & 0.12\\
     I-3 & \textbf{Agentic multiplier} $\boldsymbol{\mu_A}$ & 0.12
     \end{tabular*}}};
  \node[box=cCarbon, text width=3.85cm, align=left, anchor=north east] (co)
    at ([yshift=-6.5mm]in.south east)
    {\textbf{Carbon} \hfill \mbox{$\sum w = 0.36$}\\[1pt]
     {\scriptsize\begin{tabular*}{3.85cm}{@{}l@{\hspace{3pt}}l@{\extracolsep{\fill}}r@{}}
     I-4 & Operational carbon intensity & 0.12\\
     I-5 & Lifecycle carbon ratio & 0.12\\
     I-6 & Renewable energy fraction & 0.12
     \end{tabular*}}};
  \node[box=cWater, text width=3.85cm, align=left, anchor=north west] (wa)
    at ([yshift=-2mm]en.south west)
    {\textbf{Water} \hfill \mbox{$\sum w = 0.20$}\\[1pt]
     {\scriptsize\begin{tabular*}{3.85cm}{@{}l@{\hspace{3pt}}l@{\extracolsep{\fill}}r@{}}
     I-7 & Direct WUE & 0.10\\
     I-8 & Indirect water intensity & 0.10
     \end{tabular*}}};
  \node[box=cMaterial, text width=3.85cm, align=left, anchor=north east] (ew)
    at ([yshift=-2mm]co.south east)
    {\textbf{E-waste} \hfill \mbox{$\sum w = 0.08$}\\[1pt]
     {\scriptsize\begin{tabular*}{3.85cm}{@{}l@{\hspace{3pt}}l@{\extracolsep{\fill}}r@{}}
     I-9 & Hardware lifecycle score & 0.08\\
     \multicolumn{3}{@{}l@{}}{\strut}
     \end{tabular*}}};
  \begin{scope}[on background layer]
    \node[draw=cCarbon!75, fill=cCarbon!4, rounded corners=2pt, line width=0.6pt,
          fit=(en)(co)(wa)(ew), inner sep=3pt] (grp) {};
  \end{scope}
  \node[plainbox, text width=8.1cm, below=5mm of grp] (norm)
    {\textbf{Normalization:} each \mbox{$\hat{I}_i$} is scaled to \mbox{$[0,1]$} against a reference system (\mbox{0 = worst}, \mbox{1 = best}); I-6 and I-9 are higher-is-better, all others lower-is-better};
  \node[plainbox, text width=8.1cm, below=4mm of norm] (score)
    {\textbf{Weighted score:} \mbox{$S_{\text{SAFIA}} = \sum_{i=1}^{9} w_i\,\hat{I}_i$} with \mbox{$\sum_{i} w_i = 1$}, Eq.~\eqref{eq:safia_score}};
  \node[plainbox, text width=8.1cm, below=4mm of score] (out)
    {\textbf{Outputs:} per-model score, temporal tracking, benchmark comparison};
  \draw[arr] (in.south) -- (grp.north);
  \draw[arr] (grp.south) -- (norm.north);
  \draw[arr] (norm.south) -- (score.north);
  \draw[arr] (score.south) -- (out.north);
\end{tikzpicture}
\caption{SAFIA structure: the inputs feed nine indicators grouped by environmental dimension with their default weights, which are normalized and aggregated into the score of Eq.~\eqref{eq:safia_score}. Indicator I-3 captures the agentic multiplier $\mu_A$ of Eq.~\eqref{eq:agentic_multiplier}.}
\label{fig:safia_framework}
\end{figure}
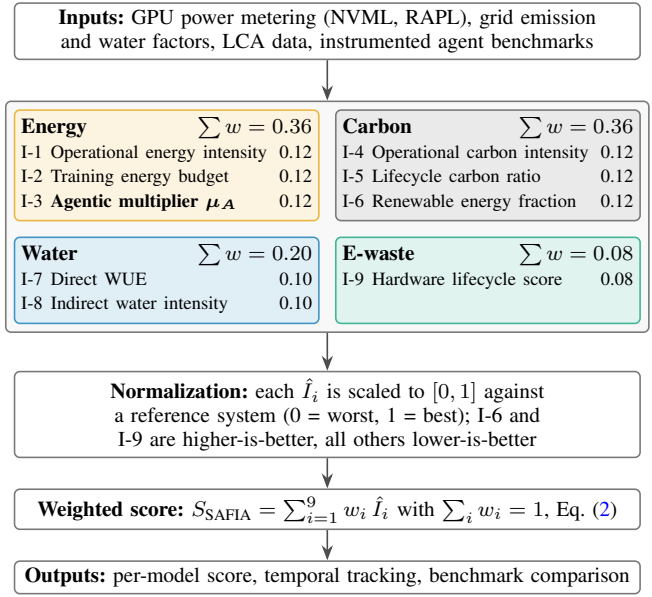

The nine indicators are aggregated into a single score for a given AI deployment:

\begin{equation}
S_{\text{SAFIA}} = \sum_{i=1}^{9} w_i \cdot \hat{I}_i
\label{eq:safia_score}
\end{equation}

where $\hat{I}_i$ is the normalized score of indicator $i$, scaled to $[0,1]$ relative to a reference system, with 0 denoting the worst and 1 the best value. I-6 and I-9 are higher-is-better indicators, whereas all other indicators are lower-is-better, so their normalization is inverted accordingly. The weight $w_i$ reflects the relative environmental priority of each dimension. Default weights are $w_1 = w_2 = w_3 = 0.12$ (energy), $w_4 = w_5 = w_6 = 0.12$ (carbon), $w_7 = w_8 = 0.10$ (water), and $w_9 = 0.08$ (e-waste), so that $\sum_{i} w_i = 1$. Practitioners can adjust these weights to reflect local environmental priorities. For example, a deployment in a water-stressed region could raise the water weight to 0.32 and lower energy and carbon to 0.30 each, keeping e-waste at 0.08 so that the weights still sum to 1. Because weights embody value judgments, we recommend reporting $S_{\text{SAFIA}}$ under both the default and the local weight vectors, together with the unweighted indicator values; if the ranking of alternative systems changes between the two, the weighting choice is decisive and should be justified explicitly. Figure~\ref{fig:safia_framework} summarizes the framework. Applying SAFIA in practice, however, depends on data that are often unavailable today, which points to the open challenges discussed next.

\section{Open Challenges and Policy Implications}
\label{sec:challenges_policy}

The analysis and the proposed framework expose gaps that research and policy must close. This section first summarizes seven open challenges and organizes them into a research roadmap, and then derives policy recommendations from them.

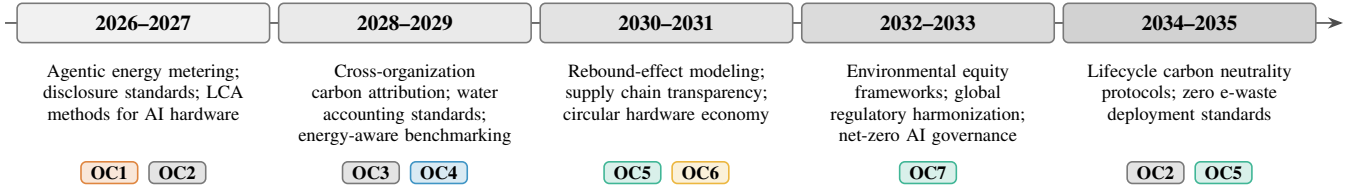
\begin{figure*}[!t]
\centering
\begin{tikzpicture}[fig]
  \def\sw{3.46}  
  \draw[arr] (-0.1,0.40) -- ({5*\sw+0.3},0.40);
  \foreach \yrs/\tint/\prio [count=\k from 0] in {
      2026--2027/8/{Agentic energy metering; disclosure standards; LCA methods for AI hardware},
      2028--2029/12/{Cross-organization carbon attribution; water accounting standards; energy-aware benchmarking},
      2030--2031/16/{Rebound-effect modeling; supply chain transparency; circular hardware economy},
      2032--2033/20/{Environmental equity frameworks; global regulatory harmonization; net-zero AI governance},
      2034--2035/25/{Lifecycle carbon neutrality protocols; zero e-waste deployment standards}} {
    \filldraw[draw=cCarbon!75, fill=cCarbon!\tint, rounded corners=2pt, line width=0.6pt]
      ({\k*\sw+0.06},0.14) rectangle ({(\k+1)*\sw-0.06},0.66);
    \node[font=\footnotesize\bfseries] at ({(\k+0.5)*\sw},0.40) {\yrs};
    \node[font=\scriptsize, text width=3.15cm, align=center, anchor=north] at ({(\k+0.5)*\sw},-0.02) {\prio};
  }
  \node[chip=cAgent]    at ({0.5*\sw-0.45},-1.58) {OC1};
  \node[chip=cCarbon]   at ({0.5*\sw+0.45},-1.58) {OC2};
  \node[chip=cCarbon]   at ({1.5*\sw-0.45},-1.58) {OC3};
  \node[chip=cWater]    at ({1.5*\sw+0.45},-1.58) {OC4};
  \node[chip=cMaterial] at ({2.5*\sw-0.45},-1.58) {OC5};
  \node[chip=cEnergy]   at ({2.5*\sw+0.45},-1.58) {OC6};
  \node[chip=cMaterial] at ({3.5*\sw},-1.58) {OC7};
  \node[chip=cCarbon]   at ({4.5*\sw-0.45},-1.58) {OC2};
  \node[chip=cMaterial] at ({4.5*\sw+0.45},-1.58) {OC5};
\end{tikzpicture}
\caption{Research roadmap to 2035 in five non-overlapping horizons.}
\label{fig:roadmap}
\end{figure*}

\begin{table*}[!t]
\caption{Policy Recommendations for AI Environmental Sustainability}
\label{tab:policy}
\centering
\footnotesize
\begin{tabularx}{\textwidth}{@{}>{\raggedright\arraybackslash}p{2.35cm} l X >{\raggedright\arraybackslash}p{3.7cm} l@{}}
\toprule
\textbf{Actor} & \textbf{Horizon} & \textbf{Recommendation} & \textbf{Mechanism} & \textbf{Priority} \\
\midrule
Regulators (EU and United States) & 2026--2027 & Mandate energy and carbon disclosure for large-scale AI training runs ($>$10$^{23}$ FLOP) and deployments ($>$10 million queries per day) & Binding regulatory requirement & High \\
Regulators & 2028--2029 & Establish an AI-specific LCA methodology standard aligned with ISO 14040/14044 & Standards body process (ISO, IEC, NIST) & High \\
Cloud providers & 2026--2027 & Provide per-workload Scope~2 energy and carbon reporting via APIs for all AI compute customers & Voluntary, then mandatory policy & High \\
Cloud providers & 2028--2029 & Achieve real-time (hourly) renewable energy matching for AI data centers & PPA restructuring and grid investment & Medium \\
Hardware manufacturers & 2028--2029 & Disclose per-chip manufacturing carbon, water, and material inventories via product environmental data sheets & Industry standardization (e.g., JEDEC) & Medium \\
AI developers & 2026--2027 & Report training compute, energy consumption, and carbon emissions in all model publications and documentation & Community norms and journal requirements & High \\
AI developers & 2026--2027 & Adopt SAFIA or an equivalent assessment framework for new model releases & Self-regulation and conference requirements & Medium \\
AI developers & 2028--2029 & Implement carbon-aware scheduling for training workloads ($>$100 kWh) & Carbon-aware scheduling software & Medium \\
Research funders & 2026--2027 & Require energy and carbon reporting in the deliverables of funded AI research & Funding condition & Medium \\
Standards bodies & 2028--2029 & Develop a standardized WUE and water accounting methodology specifically for AI workloads & ISO/IEC working group & Medium \\
\bottomrule
\end{tabularx}
\end{table*}

\subsection{Open Challenges and Research Roadmap}
\label{subsec:challenges}

Our analysis identifies seven open challenges (OC1 to OC7) in understanding and mitigating the environmental impact of AI systems. Figure~\ref{fig:roadmap} places them on a research roadmap to 2035.

\textbf{OC1: Agentic Energy Measurement Infrastructure.}
The asynchronous, multi-model, multi-tool nature of agentic pipelines defeats existing measurement tools designed for single-model training and inference. Developing agent-aware energy profiling tools that can track energy across distributed, asynchronous computation graphs is the most urgent technical research priority identified in this review.

\textbf{OC2: Embodied Carbon Attribution.}
Current AI carbon accounting overwhelmingly focuses on Scope~2 operational emissions and thus omits a substantial and sometimes dominant share of lifecycle carbon. Standardized methods for attributing GPU and server manufacturing carbon to AI workloads, analogous to capital depreciation in financial accounting, require cooperation between AI laboratories, hardware manufacturers, and foundries to establish emissions disclosure standards that are currently absent.

\textbf{OC3: Cross-Organization Emissions Attribution in Multi-Cloud.}
Modern AI deployments span multiple cloud providers, geographic regions, and organizational boundaries. Scope-based corporate accounting provides limited guidance for multi-organization AI value chains, and new attribution methodologies are needed to allocate emissions consistently across an AI deployment stack, from the foundation model trainer to the end-user application.

\textbf{OC4: Water Footprint Standardization.}
WUE metrics vary substantially in scope and methodology across data center operators. Indirect water from electricity generation, which may exceed direct cooling water in many grids, is almost universally excluded from reported figures. Standardizing water accounting to include Scope~2 and Scope~3 water is essential for accurate environmental assessment.

\textbf{OC5: Hardware Supply Chain Transparency.}
The opacity of the AI hardware supply chain, from rare earth mining through semiconductor fabrication, prevents accurate LCA of AI systems. Chip manufacturers do not publicly disclose per-chip manufacturing carbon, water, or material inventories, which makes independent LCA very difficult. Regulatory requirements for supplier sustainability disclosures would substantially improve this situation.

\textbf{OC6: Environmental Rebound Effects.}
As AI systems become more capable and energy-efficient per task, demand for AI services may grow superlinearly. If efficiency gains are outpaced by growth in usage, net environmental impact increases despite efficiency progress. Empirical study of the demand elasticity of AI services with respect to capability and cost improvements is necessary to predict and manage rebound effects.

\textbf{OC7: Environmental Equity and Global Distribution.}
AI data centers are concentrated in high-income countries with developed electrical grids and environmental regulations, whereas e-waste disproportionately affects low-income countries through informal recycling. The environmental justice dimensions of AI, namely who bears the costs and who captures the benefits, are almost entirely absent from the technical literature and require interdisciplinary research bridging environmental science, AI ethics, and development economics.

\subsection{Policy Recommendations}
\label{subsec:policy}

Several of these challenges, notably those concerning disclosure and supply chain transparency, cannot be resolved by research alone. The regulatory landscape for AI environmental sustainability remains nascent and fragmented, and environmental reporting by AI companies is therefore largely voluntary, with inconsistent methodologies, limited third-party verification, and no standardized disclosure format, which makes independent assessment and cross-company comparison unreliable. Table~\ref{tab:policy} presents our policy recommendations, organized by actor category, target horizon, and implementation mechanism.

A critical near-term priority is the disclosure of energy use for large-scale training runs. We propose a disclosure threshold of $10^{23}$ FLOP of training compute, above which pre-training energy disclosure would be mandatory; this value is our own proposal rather than an existing regulatory threshold. It is computable from training logs, does not require disclosure of proprietary model architectures, and targets the largest training runs while excluding the large volume of smaller research experiments. For deployed AI systems, we likewise propose, as our own threshold, disclosure requirements for services processing more than 10 million queries per day, with quarterly reporting of total Scope~2 energy consumption, carbon emissions, and PUE. This threshold is intended to capture large public-facing services while excluding the long tail of smaller applications. Methodological consistency should be enforced through a standardized reporting protocol, analogous to the GRI Standards for corporate sustainability reporting, developed through a multi-stakeholder process involving AI developers, regulators, standards bodies, and civil society.

\subsection{Limitations of This Review}
\label{subsec:limitations}

This work has several limitations. First, it is a conceptual synthesis of published estimates that differ in methodology, system boundaries, hardware, and grid assumptions, so these values are not directly comparable and the comparative levels are qualitative. Second, limited provider disclosure restricts the evidence available for frontier models and, especially, for agentic workloads, whose characterization is extrapolated from single-model measurements. Third, SAFIA and the agentic multiplier have not yet been applied to a deployed system; their empirical validation, including a case study and a sensitivity analysis on measured data, is left to future work. Finally, the review is restricted to English-language publications indexed in the selected databases and does not report database-specific query syntax, inter-reviewer agreement statistics, or a formal risk-of-bias assessment.

\section{Conclusion}
\label{sec:conclusion}

This review has synthesized the literature on the environmental impact of generative and agentic AI across energy, carbon, water, and e-waste. Four findings stand out. Training large models requires substantial energy, but inference can rival or exceed training over a deployment lifetime; agentic workflows multiply inference energy by one to several orders of magnitude depending on the number of model and tool calls; indirect water use and embodied carbon remain systematically underreported; and measurement, disclosure, and regulation lag behind deployment. To address these gaps, the review contributes a lifecycle taxonomy crossing lifecycle phases with five impact dimensions and GHG scopes, an analysis of agentic AI formalized through the agentic energy multiplier, the nine-indicator SAFIA framework, a comparative analysis of traditional, generative, and agentic AI, seven open challenges organized into a roadmap to 2035, and policy recommendations for regulators, cloud providers, hardware manufacturers, and AI developers. This work is a conceptual analysis that relies on heterogeneous published estimates rather than original measurements, and its conclusions are constrained by the limited disclosure of model and cloud providers. Future work should prioritize agent-aware energy instrumentation, standardized lifecycle accounting of embodied carbon and water, empirical studies of rebound effects and environmental equity, and mandatory disclosure for large-scale training and deployment, so that growth in AI capability can be decoupled from growth in its environmental footprint.

\section*{Author Contributions}
\textbf{Abderaouf Bahi}: Writing – original draft,  Methodology, Investigation, Conceptualization. 
\textbf{Amel Ourici}: Methodology, Conceptualization, Visualization, Supervision.
\textbf{Ibtissem Gasmi}: Writing – review and editing, Resources, Supervision. 

\section*{Ethical approval}
Not Applicable.

\section*{Funding}
This research received no external funding.

\section*{Declaration of competing interest}
The authors declare that they have no known competing financial interests or personal relationships that could have appeared to influence the work reported in this paper.

\section*{Acknowledgment}
The authors acknowledge the Algerian Ministry of Higher Education and Scientific Research (MESRS).

\section*{Data availability}
Not Applicable.

\section*{Declaration of Generative AI and AI-assisted technologies in the writing process}
The authors used Claude (Opus 5) to assist with drafting and language refinement. All outputs were reviewed and edited by the authors, who take full responsibility for the final content.

\bibliographystyle{IEEEtran}
\begingroup
\catcode`\&=12
\bibliography{references}
\endgroup

\end{document}